\documentclass[10pt,letterpaper,twocolumn]{article} 
\usepackage{ol}
\usepackage{hyperref}
\usepackage{amsmath,bm}
\usepackage{graphics}
\begin{document}
\twocolumn[ 
\title{Demonstration of 3-port grating phase relations}
\author{A. Bunkowski, O. Burmeister, K. Danzmann, and R. Schnabel}
\address{Albert-Einstein-Institut Hannover, Max-Planck-Institut f\"ur
Gravitationsphysik and Universit\"at Hannover, \\Callinstr. 38, 30167
Hannover, Germany}
\author{T. Clausnitzer, E.-B. Kley, and A. T\"unnermann}
\address{Institut f\"ur Angewandte Physik,
Friedrich-Schiller-Universit\"at Jena, Max-Wien-Platz 1, 07743
Jena, Germany }
\begin{abstract}
We experimentally demonstrate the phase relations of 3-port
gratings by investigating 3-port coupled Fabry-Perot cavities. Two
different gratings which have the same 1st order diffraction
efficiency but differ substantially in their 2nd order diffraction
efficiency have been designed and manufactured. Using the gratings
as couplers to Fabry-Perot cavities we could validate the results of
an earlier theoretical description of the phases at a three port
grating (\emph{Opt. Lett.} 30, p. 1183).
\end{abstract}
] 
\noindent Conventional interferometers rely on splitting and
recombining optical fields with partly transmissive beam
splitters.
When transmission through optical substrates is disadvantageous,
diffractive reflection gratings can also serve as beam splitters
allowing for all-reflective interferometry\cite{Sun97}.
As long as the grating splits an incoming beam into two outgoing
beams the phase relation at the grating and hence the properties
of the interferometer built thereof are analogous to the well
known ones of a transmissive 2-port beam splitter.
If, however, a diffractive beam splitter has more than two orders,
the mirror analog and thus the simple phase relation no longer
hold.
Yet, a knowledge of these relations at the diffractive beam
splitter is the essential premise for an understanding of multiple
port interferometry.
In a recent experiment a grating in 2nd order Littrow mount was
used to couple light into a Fabry-Perot cavity\cite{Bunkowski}.
In this case the incoming beam was split into three outgoing
beams.
The phase relations at the so-called three-port grating were
analyzed theoretically and the input-output relations for a
Fabry-Perot cavity with a three-port coupler were
derived\cite{Bunkowski05}.
The theoretical investigation of the phases was solely based on
energy conservation and reciprocity of the device but an
experimental validation of the results has not yet occurred.

%
%

In this letter we report on an experiment that was performed to
demonstrate the phase relations of optical 3-port devices.
Two different gratings were designed and manufactured for this
purpose, and used as couplers to Fabry-Perot interferometers.

%
%

Phase relations for  3-port gratings with equal diffraction
efficiencies in the $\pm$ 1st orders can be written
as\cite{Bunkowski05,Schnabel}
\begin{eqnarray}
\label{eq:phase1}
\phi_0&\!=&\!0\,,\\
\label{eq:phase2}
\phi_1&\!=&\!-(1/2) \arccos [ (\eta_1^2 - 2\eta_0^2)/(2\rho_0\eta_0)]\,,\\ 
\phi_2&\!=&\!\arccos[-\eta_1^2/(2\eta_2\eta_0)]\,,\label{eq:phase3}
\end{eqnarray}
where $\phi_0$, $\phi_1$, and $\phi_2$ are the phase shifts for
0th, 1st, and 2nd diffraction orders respectively.
\begin{figure}[htb]
    \centerline{\includegraphics[width=8cm]{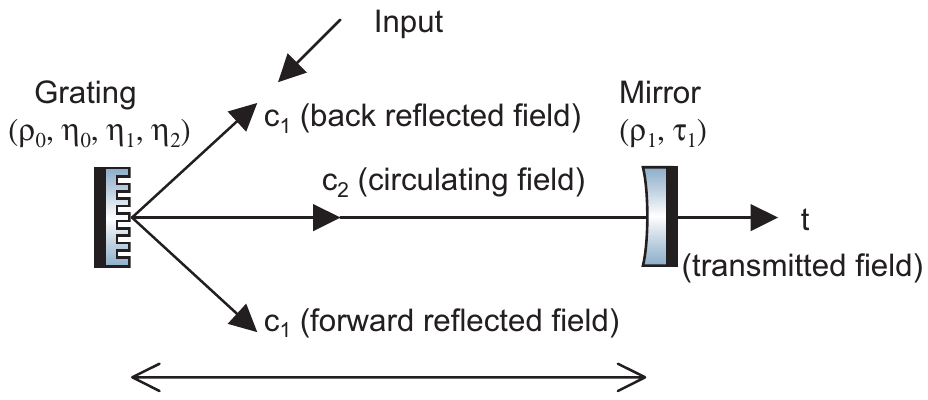}}
    \caption{(Color online) Grating in 2nd order Littrow mount
    with naming convention given in the text.} \label{fig:names}
\end{figure}
Interestingly, the coupling phases depend on the coupling
amplitudes which are given by $\eta_0, \eta_1,$ and $\eta_2$,
again,  for the 0th, 1st, and 2nd  diffraction orders
respectively, and by $\rho_0$ for the normal incidence
reflectivity of the grating.

Direct measurements of beam splitter phase relations are difficult.
If, however, the 3-port beam splitter is used to couple light into a cavity,
the cavity properties can be used to validate them.
Fig.~\ref{fig:names} shows the optical layout of a Fabry-Perot
interferometer with a 3-port grating coupler.
The grating is used in 2nd order Littrow mount and light from a
laser source is coupled to the interferometer via the grating's
1st order.
The field amplitudes of the back reflected light $(c_1)$ and
forward reflected light ($c_3$) result from interference of the
input field with the intra-cavity field and directly depend on the
phase relations between the grating ports.
In Ref.\,\cite{Bunkowski05} amplitude reflection coefficients for
 $c_1$ and $c_3$ as well as the amplitudes for the intra-cavity
field $(c_2)$ and the transmitted field $(t)$ were derived and are
repeated here for convenience:
\begin{eqnarray}\label{eq:amplitude c1}
c_1&=& \eta_2\exp(i\phi_2) + \eta_1^2\exp[2i(\phi_1+\phi)]d\,,\\
c_2&=& \eta_1 \exp(i\phi_1)d\,,\\
c_3&=&\eta_0 +
\eta_1^2\exp[2i(\phi_1+\phi)]d\,,  \label{eq:amplitude_c3}\\
 t&=&i\tau_1 c_2 \exp(i\phi)\,, \label{eq:amplitude t}
\end{eqnarray}
where the amplitude reflectance and transmittance of the cavity
end mirror are given by $\rho_1$ and $\tau_1$ respectively.
The resonance factor is given by $d
=[1-\rho_0\rho_1\exp(2i\phi)]^{-1}$ and the length $L$ of the
cavity is expressed by the tuning parameter $\phi=\omega L/c$,
where $\omega$ is the angular frequency and $c$ the speed of
light.

One distinct feature of this type of grating cavity is that the
grating phase relations allow for reflection coefficients (as a
function of $\phi$) that are not symmetric to the detuning of the
cavity.
\begin{figure}[htb]
    \centerline{\includegraphics[width=8cm]{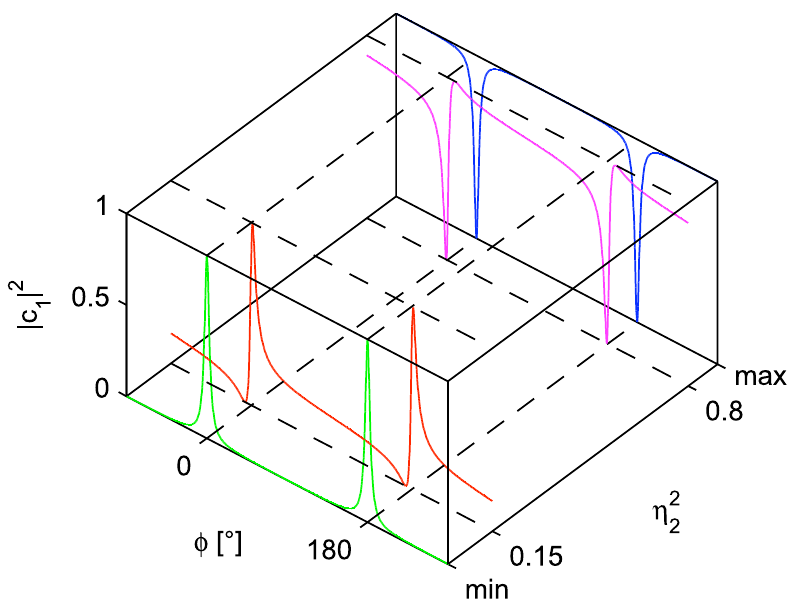}}
    \caption{(Color online) Calculated power back reflectance $|c_1|^2$ for a
    cavity with coupling $\eta_1^2=0.1$ and an end mirror with
    $\rho_1=1$ as a function of cavity tuning $\phi$ for selected values of 2nd order
    diffraction efficiency $\eta_2^2$.} \label{fig2}
\end{figure}
Fig.~\ref{fig2}  shows the calculated power back reflectance
$|c_1|^2$ of a cavity with input coupling of $\eta_1^2=0.1$ and an
ideal end mirror $(\rho_1=1)$ as a function of cavity tuning
$\phi$ for selected values of the second order diffraction
efficiency $\eta_2^2$.
In all cases shown the cavity finesse is the same. For an ideal
(lossless) grating the finesse depends on the 1st order
diffraction efficiency $\eta_1=[(1-\rho_0)/2]^{1/2}$ only.
For the minimal 2nd order diffraction efficiency \cite{Bunkowski05}
$\eta_{2,\mathrm{min}}=(1-\rho_0)/2$ all the light is reflected
back towards the laser source if the cavity is on resonance $(\phi=0
\,\,\,\mathrm{mod}\,\,\, \pi)$.
However, for maximal 2nd order diffraction efficiency
$\eta_{2,\mathrm{max}}=(1+\rho_0)/2$ no light is reflected back
from a resonating cavity.
Hence for the extremal values of $\eta_2$ the back reflected port
behaves either exactly like the reflection port or the
transmission port of a conventional impedance matched two mirror
Fabry-Perot cavity.
For intermediate values of $\eta_2$ the power reflectance is no
longer symmetric to the $\phi=0$ axis and the resonance peaks are
not of the usual Airy form as can be seen for the two exemplary
curves  $\eta_2^2= 0.15$ and $\eta_2^2=0.8$, in Fig.~\ref{fig2}.

\begin{figure}[htb]
    \centerline{\includegraphics[width=8cm]{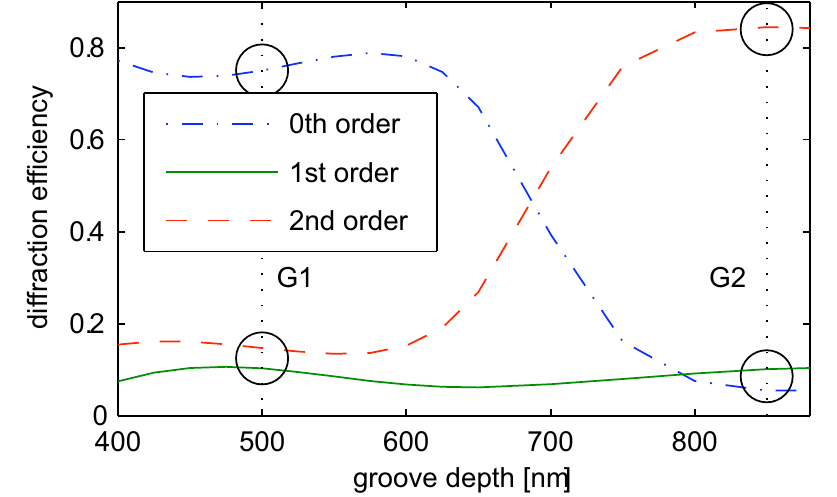}}
    \caption{(Color online) Calculated diffraction efficiencies
    as a function of groove depth obtained with RCW calculations for the gratings used.
    The circles show the design values of our gratings G1 and G2 respectively.
    }
    \label{fig:calc}
\end{figure}

To verify the grating behavior, two gratings with essentially the
same 1st order diffraction efficiency but substantially different
2nd and hence 0th order diffraction efficiency were designed and
manufactured.
The gratings use a binary structure written into the top layer of
a dielectric multilayer stack consisting of Ta$_2$O$_5$ and
SiO$_2$ placed on a fused silica substrate.
We chose a grating period of 1450\,nm  which corresponds to a 2nd
order Littrow angle of $47.2^{\circ}$ for the Nd:YAG laser
wavelength of $p=1064$\,nm used.
A rigorous coupled wave analysis\cite{Moharam} was performed to
design the grating.
The ridge width is $p/2$ and the top layer consists of 880\,nm of
SiO$_2$.
Fig.~\ref{fig:calc} shows the calculated diffraction efficiencies
for all three diffraction orders in 2nd order Littrow mount as a
function of groove depth.
The gratings were produced by ultrafast high-accuracy electron
beam direct writing~\cite{Kley} (electron beam writer ZBA23h from
Leica Microsystems Jena GmbH) and etched  by means of reactive ion
beam etching.
The etching process was stopped after reaching a groove depth of
500\,nm (G1) and 850\,nm (G2) respectively.

\begin{figure}[htb]
    \centerline{\includegraphics[width=6.5cm]{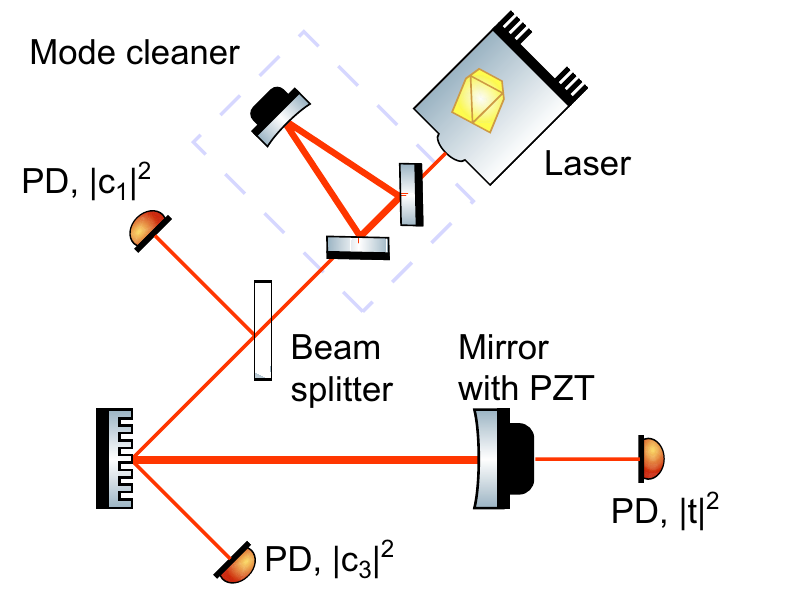}}
    \caption{(Color online) Experimental setup; PZT: piezoelectric transducer; PD: photo detector.}\label{fig:setup}
\end{figure}

A sketch of the experimental setup used to verify the grating
phase relations is shown in Fig.~\ref{fig:setup}.
A beam of a diode pumped Nd:YAG non planar ring oscillator (Model
Mephisto from Innolight GmbH) was spatially filtered with a
triangular ring cavity.
The grating (either G1 or G2) was illuminated at 2nd order Littrow
angle and a cavity end mirror with $\tau_1^2=300$\,ppm was placed
parallel to the grating's surface.
The cavity length could be controlled by a piezoelectric
transducer (PZT) and the three ports of interest were monitored by
photodetectors.

\begin{figure}[htb]
    \centerline{\includegraphics[width=8cm]{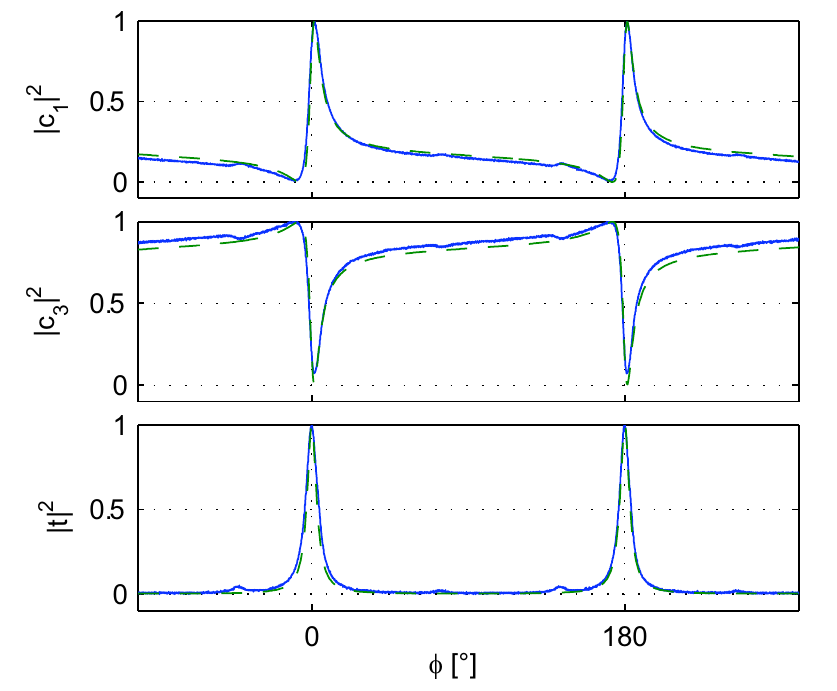}}
    \caption{(Color online) Normalized powers at the three photo detectors
    for 3-port coupler G1 as the cavity length was linearly scanned (solid, blue line)
    and the calculated values (dashed-dotted, green line).} \label{fig:gr1}
\end{figure}
\begin{figure}[htb]
    \centerline{\includegraphics[width=8cm]{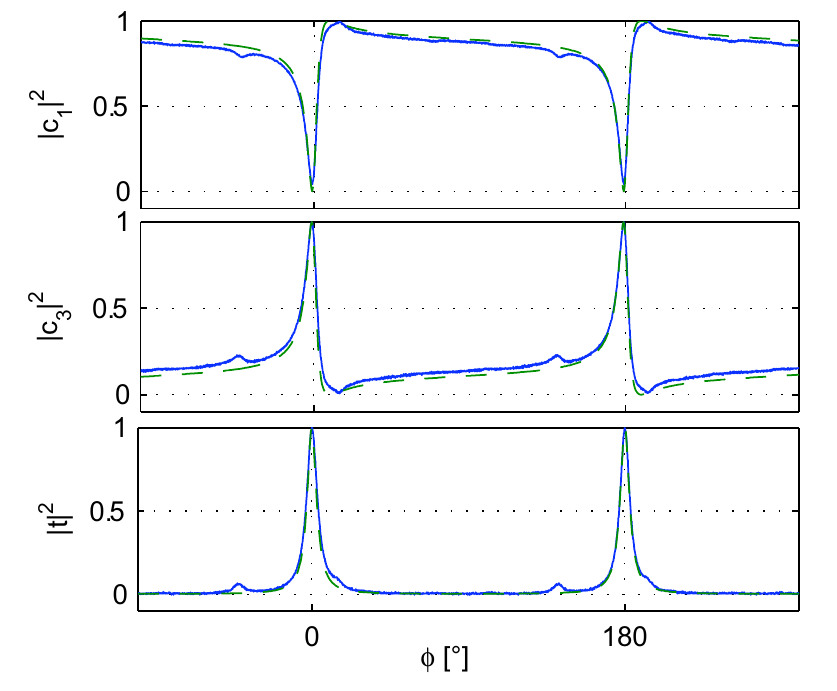}}
    \caption{(Color online) Normalized powers at the three photo detectors
    for 3-port coupler G2 as the cavity length was linearly scanned (solid, blue line)
    and the calculated values (dashed, green line).} \label{fig:gr2}
\end{figure}

Figs.~\ref{fig:gr1} and \ref{fig:gr2} show the measured signals from
the three photodetectors for linear cavity scans over one free
spectral range (FSR) using G1 and G2 respectively.
Also shown are the theoretical curves $|c_1(\phi)|^2,
|c_3(\phi)|^2,$ and $|t(\phi)|^2$ which were obtained from
Eqs.~(\ref{eq:amplitude c1}), (\ref{eq:amplitude_c3}), and
(\ref{eq:amplitude t}) using measured efficiencies of the two
gratings.
Coupling to the cavity was measured to be identical for both
gratings within the measurement accuracy of about 5\,\% of the
power meter used:
%
$\eta_1^2(\mathrm{G1}) = \eta_1^2(\mathrm{G2})=0.10$.
For the first grating  a value of  $\eta_2^2(\mathrm{G1})=0.15$ and for the second one a
value of $\eta_0^2(\mathrm{G2})=0.10$ was measured.
The remaining values were calculated using the identities
$\eta_0^2+\eta_1^2+\eta_2^2=1$ and $\rho_0^2+2\eta_1^2=1$.
We found the calculated values within the error bars of direct measurements.

Figs.~\ref{fig:gr1} and \ref{fig:gr2} show that the theoretical
and measured curves agree very well.
The interference at the 3-port gratings could therefore be well
described by the phase relations according to
Eqs.~(\ref{eq:phase1})-(\ref{eq:phase3}).
The small deviations are possibly due to imperfect mode matching,
and losses at the grating caused by transmission, scattering, and
diffraction from periodic grating errors.
As predicted, the measured intensities in the reflecting ports
showed the asymmetric behavior around cavity resonances.
%


In conclusion, we have designed and manufactured two diffraction
gratings which allowed the construction of grating-coupled
Fabry-Perot cavities with the same finesse but with totally
different properties of the two reflected ports, thereby
confirming the phase relations that were theoretically derived
earlier.
%
Our experimental results could be fully described by phase
relations based on energy conservation and reciprocity and the
knowledge of the grating's diffraction efficiencies.
No further information about the gratings was required.


This research was supported by the Deutsche Forschungsgemeinschaft
within the Sonderforschungsbereich TR7.  A.~Bunkowski's e-mail
address is alexander.bunkowski@aei.mpg.de

\end{document}